

\magnification=1200
\voffset=1.5truecm
\font\tit=cmbx12 scaled\magstep 1
\font\abst=cmsl9

\font\aun=cmbx10
\font\rauth=cmcsc10
\font\subs=cmbxti10
\topskip=1truecm
\hsize=6truein
\vsize=8.5truein
\overfullrule=0pt
\newcount\who
\who=0

\def\speak{}

\def\centra#1{\vbox{\rightskip=0pt plus1fill\leftskip=0pt plus1fill #1}}

\def\title#1{\baselineskip=20truept\parindent=0pt\centra{\tit #1}
\bigskip\baselineskip=12pt\centra{by}\def\titolo{#1}}

\def\oneauthor#1#2{\bigskip\centra{\aun #1}\medskip\centra{\it
#2}\global\def\firstauthor{#1}}


\long\def\support#1#2{\footnote{}{\hbox to 15truept{\hfill$^{#1}$\ }\sl #2}}

\parskip=5pt
\long\def\summary#1{\bigskip\centra{\speak}\bigskip\medskip
\vbox{\par\leftskip=30truept\rightskip=30truept
\noindent{\bf Summary:} \abst #1}\parindent=15truept}
\def\section#1#2{\who=1\bigskip\medskip\goodbreak{\bf\noindent\hbox to
15truept{#1.\hfil} #2}\nobreak
\medskip\nobreak\who=0}
\def\subsection#1#2{\ifnum\who=0\bigskip\goodbreak\else\smallskip\fi
{\subs\noindent\hbox to 25truept{#1.\hfil} #2}\nobreak
\ifnum\who=0\medskip\fi\nobreak}
\def\references{\bigskip\medskip\goodbreak{\bf\noindent\hskip25truept
References}\nobreak
\medskip\nobreak
\frenchspacing\pretolerance=2000\parindent=25truept}
\def\paper#1#2#3#4#5#6{\item{\hbox to 20truept{[#1]\hfill}}
{\rauth #2,} {\it #3} {\bf #4} (#5) #6\smallskip}
\def\book#1#2#3#4#5{\item{\hbox to 20truept{[#1]\hfill}} {\rauth #2,} {\it #3},
#4 (#5)\smallskip}

\newcount\firstp
\firstp=\pageno

\headline={\ifnum\pageno=\firstp\hfill\else
\line{\firstauthor,\quad{\it \titolo}\hfill\folio}\fi}
\footline={\line{\hfill}}

\title{Twistor-Like Formulation of Heterotic Strings}

\oneauthor{M. Tonin}{Dipartimento di Fisica ``G. Galilei" --
Universit\`a  di Padova}
\centerline{\it Istituto Nazionale di Fisica Nucleare -- Sezione di Padova}

\summary{
In this talk new formulations of the Green--Schwarz heterotic
strings in $D$ dimensions that involve commuting spinors,
are reviewed. These models are invariant
under $n$--extended, world sheet supersymmetry as well as under
$N=1$, target space supersymmetry where $n\leq D-2$ and
$D=3,4,6,10$. The world sheet supersymmetry replaces $n$ components
(and provides a geometrical meaning) of the $\kappa$--symmetry
in the Green--Schwarz approach. The models in $D=10$ for
$n=1,2,8$ are discussed explicitly.}

\smallskip
\noindent
(Talk given at the X Italian Conference on General Relativity and
Gravitational Physics - Bardonecchia, September 1-5 (1992)).

\smallskip

\section{1}{Introduction}

In the Neveu-Schwarz-Ramond (N.S.R.) approach [1] to
superstrings the action is invariant under super--reparametrization of
the world sheet (w.s.) i.e. it is superconformal invariant in
superconformal gauge. The covariant quantization of these models is
well understood and is consistent in ten dimensions. However the
target space (t.s.) superysmmetry is not manifest and is recovered
through the G.S.O. projection [2].
This justifies the large amount of work devoted to the alternative
Green-Schwarz (G.S.) approach [3], where the t.s. supersymmetry is
manifest. The G.S. approach, even at the classical level, can  be
formulated only in the special dimensions $D=3,4,6,10$. In the
world sheet the action is invariant under reparametrization and the
role of the w.s. supersymmetry is played here by a new local symmetry
called $\kappa$--symmetry [4]. However the geometrical meaning of
$\kappa$-symmetry in G.S. models is
rather obscure. Moreover its gauge fixing
requires an infinite chain of ghosts for ghosts (in technical terms,
$\kappa$--symmetry is infinitely reducible).

This feature, strongly
connected to the mixing of first class and second class constraints
which cannot be disentangled covariantly, is the source of the
difficulties of the covariant quantization in the G.S. approach.
$\kappa$--symmetry is also present in superparticle [5] (and
supermembrane [6]) models. A consistent covariant quantization is now
available for superparticles [7] but the problem is still open in the
case of superstrings (and supermembranes). Then it is
useful to look at superparticles and G.S. superstrings (or
supermembranes) from different points of view. In particular it has
been suggested [8]--[11] that twistors [12] could play a r$\hat o$le
here.

In an important paper, Sorokin et al. [13] have proposed a new formulation
for superparticles in $D=3,4$. Subsequently this formulation, or
related ones, have been worked out both for superparticles
[14]--[21] and G.S.
superstrings [22]--[32] in the special dimensions $D=3,4,6,10$.
As in previous approaches, these formulations involve commuting
spinors i.e. twistor-like variables. In addition, they show both
manifest target space supersymmetry and $n$-extended
world line/world sheet
supersymmetry $(1\leq n \leq D-2)$. The supersymmetry of the world
manifold
replaces $n$ components of the $\kappa$-symmetry and therefore
provides a geometrical meaning of that symmetry. At least at
the classical level, the maximal extended models, with $n=D-2$,
are of special interest since, in this case, the whole $\kappa$--symmetry
is replaced by (extended) supersymmetry of the world manifold.

The hope is that twistor models could help in solving the
problem of covariant quantization of G.S. superstrings.
Significant steps in that direction have been done by
Berkovits [31],[32].
In any case these models are useful to clarify the G.S. approach and the
geometrical meaning of the $\kappa$--symmetry.

In the next two sections I shall review briefly the G.S. approach of
the heterotic string (sect. 2) and I shall discuss
some properties
of commuting spinors (twistors)
(sect. 3). Then, in section 4, I shall describe the new
twistor--like formulations of the heterotic strings in $D=10$ with
$n=1,2$ and 8.

\section{2}{The G.S. Approach}
The heterotic string, in the G.S. approach, describes the
embedding of the two dimensional world sheet ${\cal M}(2| 0)$ on
the target superspace $\underline{{\cal M}}(D|2(D-2))$ where
$D=3,4,6,10$.

The world sheet is parametrized locally by the coordinates
$\xi^i\equiv(\xi^{(+)}, \xi^{(-)})$. The w.s. zweinbeins
$e^\pm$ have vanishing torsion, so that the
Lorentz connection can be expressed in terms of $e^\pm_i(\xi)$.
The w.s. differential is
$
d=d\xi^i\partial_i = e^+ D_+ + e^-D_-
$.

The target superspace has $D$ bosonic dimensions and $2(D-2)$
fermionic dimensions and is parametrized locally by
the w.s. scalar, string coordinates, $Z^M\equiv(X^m,\theta^\mu)$
where $X^m$ $(m=0,1,...D-1)$ are t.s. vectors and $\theta^\mu$
$(\mu=1,... 2(D-2)$ are Majorana, t.s. spinors (Weyl-Majorana in $D=10$).
The field content of the model is completed by a set of heterotic
fermions $\psi_r\ (r=1,...N)$ which are left--handed, w.s. Weyl-Majorana
spinors. In $D=10$ they are essential, at the quantum level, to
cancel the left--handed conformal anomaly (then $N=32$).

To write the action, in flat background, let us define

$$\eqalign{
{\cal E}^a_\pm & = D_\pm X^a - (D_\pm\theta\Gamma^a\theta);\quad
{\cal E}^\alpha_\pm = D_\pm\theta^\alpha\cr
B^{(0)}_{\alpha\beta} & = 0 = B^{(0)}_{ab};\quad B^{(0)}_{a
\beta} = - B^{(0)}_{\beta a} = - (\Gamma_a)_{\beta\gamma}\theta^\gamma
\cr}
$$

\noindent
where $(\Gamma^a)_{\beta\gamma}=(\gamma^a C)_{\beta\gamma}$ and
$(\Gamma^a)^{\beta\gamma}=(C^{-1}\gamma^a)^{\beta\gamma}$ are
symmetric in $\beta,\gamma$ ($\gamma^a$ are Dirac matrices and
$C$ is the charge conjugation matrix). Moreover
we shall write  ${\cal E}^A_\pm = ({\cal E}^a_\pm, {\cal E}^\alpha_\pm)$.
Then the G.S. action is

$$
I=\int e^+\wedge e^- \{{\cal E}^a_+{\cal E}_{-a} + {\cal E}^A_+
{\cal E}^B_- B^{(0)}_{BA} + i \sum^N_{r=1} (\psi_r D_-\psi_r)\}\eqno(2.1)
$$

The first term contains the kinetic action, the second is the
Wess-Zumino term and the third term is the heterotic action. One
should notice that the Virasoro condition
$
{\cal E}^a_-{\cal E}_{-a} = 0
$
is among the field equations of this action (it is obtained by
varying $e^i_+(\xi))$.

In addition  to w.s. reparametrization, Weyl and Lorentz local
transformations, the action (1) is invariant under the local
$\kappa$--symmetry:

$$\eqalign{
\delta_\kappa\theta^\alpha & = {\cal E}^a_-(\Gamma_a)^{\alpha\beta}
K_\beta\quad ; \quad \delta_\kappa X^a=(\delta_\kappa\theta\Gamma^a\theta)\cr
\delta_\kappa e^i_+ & = 4({\cal E}^\alpha_+ K_\alpha)e^i_-\quad ; \quad
\delta_\kappa e^i_- = 0 = \delta_\kappa\psi_r\cr}\eqno(2.2)
$$

\noindent
where $K_\alpha$ are the anticommuting,
$\kappa$--symmetry gauge parameters ($K_\alpha$ is a commuting
ghost in the BRS version of the $\kappa$--symmetry).

Since on shell ${\cal E}^a_-\Gamma_a$ is a projector, there are
secondary gauges that affect $K_\alpha$ and involve secondary ghosts
and so on,
so that
$\kappa$--symmetry is infinitely reducible. Then half of the $2(D-2)$
components of $K_\alpha$ are pure gauge and $\kappa$--symmetry contains
only $(D-2)$ parameters.

It is easy to extend this model in presence of a SUGRA-SYM background
[33] (G.S., $\sigma$--model). The tangent target space geometry
is described in terms of the supervielbeins
$E^A=dZ^M E^A_M(Z)$, the Lorentz superconnection $\Omega_A^{\phantom{A}B}=
dZ^M \Omega_{MA}^{\phantom{MA}B}(Z)$, the gauge superconnection
$A=dZ^M A_M(Z)$ and the two--superform\hfill\break
$B=dZ^M dZ^N B_{NM}(Z)$. The curvatures
of these superforms are respectively the torsion, $T^A$, the
Lorentz curvature, $R_A^{\phantom{A}B}$, the gauge curvature, $F$,
and the $B$--curvature, $H=dB$. The intrinsic components of the w.s.
pull--back of $E^A$ are denoted $E^A_\pm$.
The flat case is recovered in the limit
$
E^A\rightarrow{\cal E}^A\ ;\ B\rightarrow B^{(0)}\ ;\
A\rightarrow 0
$.

In these notations the action is

$$
I=\int e^+\wedge e^- \{\phi E^a_- E_{+a}+E^A_- E^B_+ B_{BA}  + i \psi
(D_--E^B_- A_B)\psi\}\eqno(2.3)
$$

\noindent
where $\psi$ is the column vector $(\psi_r)$ and $\phi(Z)$ is the
dilaton superfield.

\noindent
Now the Virasoro condition is

$$
E^a_- E_{-a}=0\eqno(2.4)
$$

\noindent
and the $\kappa$--symmetry transformations are

$$\eqalign{
\delta_\kappa Z^M E^\alpha_M & = w^\alpha;\ \delta_\kappa Z^M E^a_M=0;\
\delta_\kappa e^i_-=0\cr
\delta_\kappa e^i_+ & = 4 e^i_-(E^\alpha_+ K_\alpha)+...;\
\delta_\kappa\psi = w^\beta A_\beta\psi\cr}\eqno(2.5)
$$

\noindent
where
$
w^\alpha = E_-^a (\Gamma_a)^{\alpha\beta} K_\beta
$ and the dots represent terms involving curvature components.
However $I$ is invariant under $\kappa$--symmetry only when the
SUGRA-SYM background constraints [34],[9] are imposed:

$$\eqalign{
T^a_{\alpha\beta} & - 2 \Gamma^a_{\alpha\beta}
= 0 = T^\gamma_{\alpha\beta}\cr
H_{\alpha\beta\gamma} & = 0 = H_{a\beta\gamma} -
\phi(\Gamma_a)_{\beta\gamma}\cr
F_{\alpha\beta} & = 0 \cr}\eqno(2.6)
$$

\noindent
In D=10,
eqs.(6) are just the constraints that lead to the field equations
for the decoupled, $N=1$, Supergravity and Super Yang-Mills theory
(the coupling arises from $\sigma$--model quantum
corrections [35],[36]).

\section{3}{Twistors}
Twistors are a deep mathematical concept introduced by Penrose [12].
Roughly speaking a twistor is a couple of commuting spinors
$(\lambda^\alpha, w_\alpha)$ and a point in the twistor--space corresponds
to a light--line in space time. An ambitious program is to study
Q.F.T. (in particular quantum gravity) in twistor space $(\lambda, w)$
rather than in the usual phase space $(X,P)$.

\noindent
However we shall use the word ``twistor" in a more superficial way.
For us, a twistor (strictly speaking: an half--twistor) is simply a
commuting Majorana spinor $\lambda^\alpha$ in the special
dimensions $D=3,4,6,10$ where it has $2(D-2)$ real components.
Moreover, due to the fundamental $\Gamma$--matrix identity

$$
\Gamma^a_{\alpha\beta} \Gamma_{a\gamma\delta} + \Gamma^a_{\beta\gamma}
\Gamma_{a\alpha\delta} + \Gamma^a_{\gamma\alpha}
\Gamma_{a\beta\delta} = 0\eqno(3.1)
$$

\noindent
which holds in these (and only these) dimensions, the vector
$v^a=(\lambda\Gamma^a\lambda)$ is light--like: $v^a v_a=0$.
Therefore, if the momentum $p^a$ of a (super)particle is written as
$p^a=(\lambda\Gamma^a\lambda)$, the massless condition $p^a p_a=0$
is automatically satisfied. Similarly, in G.S. superstrings, the
constraint $E^a_-=(\lambda\Gamma^a\lambda)$ implies the Virasoro
condition $E^a_- E_{-a} =0$.
In addition, since in Majorana representation $\Gamma^0=1$, one has

$$
v^0 = (\lambda\Gamma^0\lambda)>0\eqno(3.2)
$$

\noindent
so that the light--like vector $v^a$ points in the future light--cone.

It is worthwhile to report at this point a remark of
Bengtsson et al.[10]. In first quantized models, covariance requires both
positive
and negative energy states. But supersymmetry $(\{Q,Q\}=H!)$ requires
positive energy. This is at the origin of the difficulty for the
covariant quantization of superparticle and G.S. superstring models.
Owing to eq. (2), quantization in twistor--space could be the key to
overcome this problem.

In any case, coming back to our discussion, we have shown that
twistors describe light--like vectors that point in the future
light--cone. However a light--like vector $v^a$, with $v^0>0$
and modulo a scaling, parametrizes the sphere $S^{D-2}$. In fact
one has $\sum\limits^{D-1}_{i=1} v^2_i = v^2_0$. On the other hand,
a twistor $\lambda^\alpha$, modulo a scaling, parametrizes the sphere
$S^{2(D-2)-1}$. In fact, since $\Gamma^0=1$, one has
$
\sum\limits^{2(D-2)}_{\alpha=1} (\lambda^\alpha)^2 =
v^0 > 0
$

\noindent
Then the problem is how to make the twistor description
equivalent to the light--like vector one i.e. how to get
the sphere $S^{D-2}$ parametrized by twistors.

The problem would be solved if it was possible to consider twistors
modulo gauge transformations that belong to the cosets
${S^{2(D-2)-1}\over S^{D-2}}$. These cosets are
$Z_2, S^1, S^3$ and $S^7$ respectively, in the magic dimensions
$D=3,4,6,10$ (an isomorphism known as Hopf fibration). Moreover
the Hopf fibration is related to the division algebras of real,
complex, quaternionic and octonionic numbers, i.e. R,C,H,O. Then
a simple idea is to consider twistors valued in the division
algebras R,C,H,O respectively in $D=3,4,6,10$.
However Table I shows that an obstruction has to be waited
in the most relevant dimension $D=10$. Indeed in this case
$S^7$ is not a group and the division algebra of octonions is
not associative.

\midinsert
$$
\vbox{\tabskip=0pt\offinterlineskip
\def\tablerule{\noalign{\hrule}}
\halign to 15truecm{\strut#&\vrule#\tabskip=1.em plus 1.5em&
\hfill#\hfill & \vrule#& \hfill#\hfill & \vrule#& \hfill#\hfill &
\vrule#& \hfill#\hfill & \vrule#& \hfill#\hfill &
\vrule#\tabskip=0pt\cr
\tablerule
&&   &&   &&   &&   &&    &  \cr
\noalign{\vskip -0.4truecm}
&& D && 3 && 4 && 6 && 10 &  \cr
&&   &&   &&   &&   &&    &  \cr
\noalign{\vskip -0.4truecm}
\tablerule
&&   &&   &&   &&   &&    &  \cr
\noalign{\vskip -0.4truecm}
&& 2(D-2) && 2 && 4 && 8 && 16 & \cr
&&   &&   &&   &&   &&    &  \cr
\noalign{\vskip -0.4truecm}
\tablerule
&&   &&   &&   &&   &&    &  \cr
\noalign{\vskip -0.4truecm}
&& $S^{D-2}$ && $S^1$ && $S^2$ && $S^4$ && $S^8$ & \cr
&&   &&   &&   &&   &&    &  \cr
\noalign{\vskip -0.4truecm}
\tablerule
&&   &&   &&   &&   &&    &  \cr
\noalign{\vskip -0.4truecm}
&& $S^{(D-2)-1}$ && $S^1$ && $S^3$ && $S^7$ && $S^{15}$ & \cr
&&   &&   &&   &&   &&    &  \cr
\noalign{\vskip -0.4truecm}
\tablerule
&&   &&   &&   &&   &&    &  \cr
\noalign{\vskip -0.4truecm}
&& ${S^{(D-2)-1}\over S^{D-2}}$ && $Z_2$ &&  $S^1\approx U(1)$ &&
$S^3\approx SU(2)$ && $S^7$ not\ a\ group & \cr
&&   &&   &&   &&   &&    &  \cr
\noalign{\vskip -0.4truecm}
\tablerule
&&   &&   &&   &&   &&    &  \cr
\noalign{\vskip -0.4truecm}
&& ${\rm Division\atop \rm Algebra}$ && {\bf R} && {\bf C} && {\bf H} &&
{\bf O} not\ associative & \cr
\tablerule
}}
$$
\smallskip
\centerline{Table I}
\endinsert

Recently a way has been found to overcome this $D=10$ barrier.
The method is based on the isomorphism [19],[20]

$$
SO^{D-2}\simeq
{Spin(1,D-1)\over [SO^\uparrow(1,1)\otimes S0 (D-2)]\times K}
$$

\noindent
where $Spin(1,D-1)$ is the covering of the Lorentz group in D
dimensions, $SO^\uparrow(1,1)$ is the orthochronous Lorentz group
in $D=2$, $K$ is the abelian group of the special conformal
transformations in $D-2$ dimensions (conformal boosts) and $\times$
denotes a semidirect product.
The explicit
construction of this isomorphism in terms of twistors is the following
[21].
Consider $D-2$ twistors $\lambda^\alpha_q$ $(q=1,...D-2)$ such that

\noindent
i) $(\lambda_p\Gamma^a\lambda_q)= v^a\delta_{qp}$ \hfill (3.3)

\noindent
ii) $\lambda^\alpha_q$ are defined modulo $SO^\uparrow(1,1)\otimes
   SO(D-2)$ gauge transformations;

\noindent
iii) The $(D-2)\times 2(D-2)$ matrix $\lambda^\alpha_q$ has highest
   rank: rank $||\lambda ||=D-2$.

It has been shown in [21] that $D-2$ twistors that fulfil these
conditions parametrize the sphere $S^{D-2}$.

This construction is the geometrical basis of the maximally extended
twistor--like models of superparticle and heterotic strings. These
models have $(D-2)$--extended world sheet supersymmetries and the
twistors arise as the superpartners of $\theta^\alpha$.
Of course one can construct intermediate models with a smaller
number of twistors i.e. a smaller number of w.s. supersymmetries.

\section{4}{The twistor approach to G.S. heterotic strings}
In this section I shall present the twistor--like formulations
of G.S. heterotic strings in $D$ dimensions, with $N=(n,0)$--extended,
w.s. supersymmetry and $n\leq D-2$. These models describe the
embedding of the superworld sheet ${\cal M}(2|n)$ into the target
superspace ${\underline{\cal M}}(D|2(D-2))$.
I shall consider explicitly only
the case with $D=10$ and $n=1,2,8$. The superworld sheet is parametrized
locally by the supercoordinates $\xi^I=(\xi^{(+)}, \xi^{(-)},
\eta^{(q)})$ where $\eta^{(q)}$ are real grassmann parameters
and $q=1,...,n$. The supervielbeins are $e^{\cal A}\equiv
(e^+, e^-, e^q)$ so that the differential is $d=e^+ D_+ +
e^- D_- + e^q D_q$. The w.s. tangent space structure
group is $SO^\uparrow(1,1)\otimes SO(n)$ i.e. the product of
the orthochronous $D=2$ Lorentz group and the
group $SO(n)$ that acts on the $e^qs$. We shall call $\Delta$
the covariant differential with respect to the Lorentz and the
$SO(n)$ group. The w.s. torsion $T^{\cal A} = \Delta e^{\cal A}$
is constrained as follows:

$$
T^-=\sum_q e^q \wedge e^q;\ T^+=0\ ; T^q=e^+\wedge e^-
{\cal T}^q_{-+}
$$

As in the G.S. approach, the target fields are the string
supercoordinates and the heterotic fermions but now they are w.s.
superfields:

$$
\hat Z^M(\xi^{(\pm)}, \eta^{(q)})\equiv(\hat X^m,\hat\theta^\mu);
\qquad\qquad \hat\psi(\xi^{(+)},\eta^{(q)})
\equiv(\hat\psi_r)
$$

The geometry of the target space is described in terms of the
supervielbeins $\hat E^A=E^A(\hat Z)$, the superconnections
$\hat\Omega^A_A=\Omega^B_A(\hat Z)$, $\hat A=A(\hat Z)$ and the
two-superform $\hat B=B(\hat Z)$.
The SUGRA-SYM
constraints (2.6) are imposed as w.s. superfields. The intrinsic
components of the pull back of $\hat E^A$ on the superworld sheet
are $(\hat E^A_\pm, \hat E^A_q)$ and we shall write

$$
(\hat E^A_\pm, \hat E^A_q)\Bigl\vert_{\eta^{(q)}=0}
= (E^A_\pm, \lambda^A_q)
$$

One should notice that $\lambda^\alpha_q$ are commuting spinors
(twistors).

In order to formulate twistor--string models with $(n,0)$--extended,
w.s. supersymmetry, two key ingredients are needed. The first
ingredient is to
impose a suitable constraint to enforce the fundamental
twistor relation $E^a_-=(\lambda\Gamma^a\lambda)$. Of course this
constraint must be written as a superfield equation in order to
preserve the w.s. supersymmetry. It turns out that the right
constraint is

$$
\hat E^a_q=0 \qquad\qquad (q=1,...,n)\eqno(4.1)
$$

This constraint means that the embedding of the superworld sheet
in the target superspace is such that the odd part of the tangent
superworld sheet lies entirely within the odd part of the tangent
target superspace. Eq. (1), developed in power of $\eta$, gives
the component constraints:

$$
\lambda^a_q=0\eqno(4.2)
$$

$$
(\lambda_p\Gamma^a\lambda_q) =
\delta_{pq} E^a_-\eqno(4.3)
$$

\noindent
together with further constraints, irrelevant here.

If $n=D-2$, eq. (3) is just eq. (3.3). Moreover, since here
the structure group is
$S^\uparrow(1,1)\otimes SO(D-2)$, $\lambda^\alpha_q$ transform
under gauge transformations that belong to this group. Then,
by assuming that the rectangular matrix $\lambda^\alpha_q$ has
highest rank, we meet the situation described at the end of
section 3: the twistors $\lambda^\alpha_q$ parametrize the
sphere $S^{D-2}$ and therefore describe the light--like vector $E^a_-$
with $E^0_->0$.

For models with $n=1$, eq. (1) is sufficient to write the action in a
manifestly supersymmetric form i.e. as a full superspace integral. On
the contrary if $n>1$, the action term $I^{(B)}$ that involves the
two--form $B$
cannot be written naively as a
full superspace integral without introducing new auxiliary
superfields.

The second ingredient needed to write a supersymmetric action
$I^{(B)}$, if $n>1$, is an interesting property of the two
superform $\hat B$, that holds if the twistor constraint, eq. (1),
and the SUGRA--SYM constraints (the superfield version of eq. (2.6)),
are satisfied. Indeed in this case there exists a modified
two--superform [30]

$$
\tilde B = \hat B + {1\over n}
e^+ \wedge e^- \sum_q \hat E_q^A
\hat E^B_q \hat E^C_+ \hat H_{CAB}\eqno(4.4)
$$

\noindent
such that the pull--back on the superworld sheet ${\cal M}$ of
$d\tilde B$ vanishes

$$
d\tilde B |_{\cal M} = 0\eqno(4.5)
$$

\noindent
This fact allows to write $I^{(B)}$ in two equivalent ways. \hfill\break
\noindent
\underbar{I$^0$ way:}\footnote*{This is the approach of refs. [27], [28].
However, since the author of refs. [27], [28] was not yet aware of
the Weil triviality of $\hat B$, he imposed an additional, unneccesary
constraint on $\hat B$, to recover the w.s. supersymmetry.} Due to
eq. (5), under the infinitesimal super--reparametrization
$\xi^I\rightarrow\xi^I + \epsilon^I$

$$
\delta_\epsilon \tilde B = (di_\epsilon\tilde B + i_\epsilon d\tilde B)
|_{\cal M} = d i_\epsilon \tilde B |_{\cal M}\eqno(4.16)
$$

\noindent
so that, if ${\cal M}_0$ is the slide of ${\cal M}$ at
$\eta^{(q)}=0=d\eta^{(q)}$, the action

$$
I^{(B)} = \int_{{\cal M}_0} \tilde B\eqno(4.6)
$$

\noindent
is super--reparametrization invariant, even if it is not a full
superspace integral. The same mechanism [37], called Weil triviality, is
operative in S.Y.M. theories and permits in that context to apply
the descent equation method to get the consistent chiral anomaly
of these models.

\vskip 0.5truecm

\noindent
\underbar{II$^0$ way:}
\footnote{**}{This is the approach of refs.
[21],[30], firstly proposed in a different context in ref. [38].}
Eq. (5) implies that locally $\tilde B|_{\cal M}=d\hat Q|_{\cal M}$.
Then $I^{(B)}$ can be written as a full superspace integral simply by
imposing this condition as a constraint:

$$
I^{'(B)} = \int_{\cal M} \hat P^{IJ}
(\hat B_{IJ}-\partial_I \hat Q_J)\eqno(4.7)
$$

\noindent
where the lagrangian multipliers $\hat P^{IJ}$ are
Grassmann--antisymmetric w.s. superfields. $I^{'(B)}$ is invariant under
the local gauge transformations (Bianchi gauge)

$$
\delta\hat P^{IJ} = \partial_K \hat\Lambda^{KIJ}\eqno(4.8)
$$

\noindent
where the gauge parameters $\hat\Lambda^{KIJ}$ are Grassmann--antisymmetric
in their indices. This Bianchi gauge allows to prove the equivalence
between $I^{(B)}$ and $I^{'(B)}$.

Now let us discuss more explicitly the $D=10$, $(n,0)$--extended,
twistor--string models for $n=1,2,8$.

\subsection{4.1}{D=10, N=(1,0), heterotic, twistor--string model
[22],[27]}
\noindent
Here the superworld sheet has only one odd dimension and we
shall write $(\hat E^A_\pm, \hat E^A_1)_{\eta=0}$ =
$(E^A_\pm, \lambda^A)$. Now the twistor constraint is

$$
\hat E^a_1=0
$$

\noindent
which implies

$$
\lambda^a=0;\quad\quad E^a_-=(\lambda\Gamma^a\lambda)\eqno(4.9)
$$

\noindent
As said before, the action can be written as a full superspace
integral without resort to the property of Weil triviality.

The superspace action is

$$
I=\int d^2\xi\ d\eta\ sdet\ e
\Bigl\{ \hat P_a \hat E^a_1 + \hat E^A_+ \hat E^B_1 \hat B_{BA} + i
\hat\psi {\cal D}_1 \hat\psi\Bigr\}\eqno(4.10)
$$

\noindent
where the superfield $\hat P_a=P_a+\eta\beta_a$ are lagrangian
multipliers, $\hat\psi_r=\psi_r+\eta\chi_r$ are the heterotic
fermions and

$$
{\cal D}\hat\psi = e^+{\cal D}_+\hat\psi + e^- {\cal D}_-\hat\psi
+ e^1 {\cal D}_1\hat\psi = d\hat\psi - \hat E^B\hat A_B
\hat\psi
$$

\noindent
is the covariant differential of $\hat\psi$ in presence of the
background, gauge superconnection $\hat A$. The first term in the r.h.s. of
eq. (10) imposes the twistor constraint, the second one
contains the Wess-Zumino
term and the last one is the heterotic action. In components,  eq. (10)
becomes

$$
I=\int d^2\xi\ det\ e \{P'_a(E^A_- -\lambda\Gamma^a\lambda)+
\phi E^a_+ E_{-a} + E^A_+ E^B_- B_{BA} + i \sum_r\psi_r
{\cal D}_- \psi_r\}\eqno(4.11)
$$

To get eq. (11) the SUGRA-SYM constraints have been used, the non
propagating fields $\lambda^a$ and $\chi_r$ have been eliminated through
their field equations and the shift $P_a=P'_a-\phi(\lambda\Gamma^a
\lambda)$ has been performed.

The action (10) is manifestly invariant under w.s. reparametrization,
Weyl, Lorentz and local $N=(1,0)$ supersymmetry. It is also
invariant under the two--fold reducible, residual $\kappa$--symmetry
with only 7 superfield parameters i.e. 7+7 parameters (the 8th one
having been absorbed by the w.s. supersymmetry).

These $\kappa$--transformations are

$$\eqalign{
\delta_\kappa \hat Z^M \hat E^\alpha_M & \equiv \hat w^\alpha =
(\hat E_1\Gamma^a\hat E_1)\Gamma^{\alpha\beta}_a \hat K_\beta -2
(\hat E_1 \hat K) \hat E^\alpha_1;\cr
\delta_\kappa \hat Z^M \hat E^a_M & = 0;\ \delta_\kappa \hat P^a = 4i
(\hat w \Gamma^a \hat E_+);\ \delta_\kappa\hat\psi=\hat w^\alpha
\hat A_\alpha\hat\psi\cr}
$$

Moreover the action (10) is also invariant under the local
symmetry ($\beta$--symmetry)

$$
\delta P^{'a}=\beta(E^a_-+\lambda\Gamma^a\lambda);\quad
\delta e_+=-\beta e^i_-
$$

{}From this symmetry one can show that the field equations from the
action (11) are the same as in the G.S. approach. Indeed,
by varying $\lambda^\alpha$ one get $P'_a\Gamma^a\lambda=0$ so that

$$
P^{'a} = \gamma (\lambda\Gamma^a\lambda)= {1\over 2}\gamma
(E^a_-+\lambda\Gamma^a\lambda)
$$

\noindent
and one can choose $\gamma=0$ by gauge fixing the $\beta$--symmetry.
This (1,0)--model does not seems suitable for
quantization, or, to say better, the natural quantum version of this
model is the...N.S.R. string. To be more explicit, let us
consider the model in
flat target space and let us impose the following, non
covariant gauge fixing for the (super) $\kappa$--symmetry:

$$
\theta^\alpha-Q^a \Gamma^{\alpha\beta}_a N_\beta=0\eqno(4.12)
$$

$$
\lambda^\alpha - \Lambda^a \Gamma_a^{\alpha\beta} N_\beta=0\eqno(4.13)
$$

\noindent
where $N_\alpha$ is a constant, $D=10$, W--M spinor and
the new fields $\Lambda^a, Q^a$ are restricted by the conditions
$\Lambda^a\Lambda_a=0=\Lambda^a Q_a$.
Then $\Lambda^a$ and $Q^a$ have 9
components each so that eqs. (12), (13) correspond to 7+7
conditions. If this gauge is imposed partially to fix only the
superpartners of $K_\alpha$ and one integrate over $\lambda^\alpha,
\Lambda^a$ and lagrangian multipliers, one gets formally the G.S.
string action.

But now let us impose the full gauge fixing (12), (13) then integrate
over $\theta^\alpha$, $\lambda^\alpha$, $\Lambda^a$ and Lagrangian
multipliers and perform the change of variables [29]

$$\eqalign{
\bar X^a & =X^a + i Q^a(n\cdot(Q)\cr
\bar\phi^a & = 2\sqrt{n\cdot{\cal E}_-)}
\Bigl[ Q^a + {1\over 2} {(n\cdot Q)(Q\cdot D_- Q)\over
(n\cdot{\cal E}_-)}\Bigr]\cr}
$$

\noindent
where
$
n^a =(N\Gamma^a N)
$.
Then from eq. (10) one recovers the action for the N.S.R. string.
Two comments are in order:

\smallskip
\noindent
i) Since here the $\kappa$--symmetry is (two-fold) reducible, one
must fix also the secondary gauges. If this is done properly, the
factors $(n\cdot{\cal E}_-)$ that arise from the path integrations
cancel exactly.

\smallskip
\noindent
ii) Since from eq. (9), ${\cal E}^0_->0$,
the G.S. and N.S.R. models obtained through the procedure
outlined above, have their phase spaces halved with respect
to those of the usual formulations. For further details see ref. [29].

\subsection{4.2}{D=10, N=(2,0), heterotic, twistor--string model
[27],[31],[32]}

Since now the structure group is $SO^\uparrow(1,1)\otimes SO(2)$
and $SO(2)$ is isomorphic to $U(1)$, one can choose a single complex,
grassmann parameter $\eta$ (with its conjugate $\bar\eta$) as coordinate
of the odd dimensions of the superworld sheet. As for the w.s. pull back
of the target supervielbeins $\hat E^A$, we shall write

$$
(\hat E^A_\pm, \hat E^A_1, \hat{\bar E}^A_1)_{\eta=0} =
(E^A_\pm, \lambda^A, \bar\lambda^A).
$$

$\lambda^\alpha$ are complex, commuting, t.s. Weyl--Majorana spinors.

Now the twistor constraint, in superfield language, is

$$
\hat E^a_1 = 0 = \hat{\bar E}^a_1
$$

In components, it yields

$$
\lambda^a = 0 = \bar\lambda^a
$$

$$
(\lambda\Gamma^a\lambda) = 0 = (\bar\lambda\Gamma^a\bar\lambda)
\eqno(4.14)
$$

$$
E^a_- = (\lambda\Gamma^a\bar\lambda)\eqno(4.15)
$$

A complex, commuting spinor $\lambda^\alpha$
in  D=10 that satisfies eq. (4.14) is
called a \underbar{pure spinor}. It is equivalent to a couple of
twistors: $\lambda^\alpha = \lambda^\alpha_1 + i\lambda^\alpha_2$
such that

$$
(\lambda_1,\Gamma^a\lambda_1) = (\lambda_2\Gamma^a\lambda_2)\quad ;\quad
(\lambda_1\Gamma^a\lambda_2) =0
$$

For a pure spinor, $(\lambda\Gamma^a\bar\lambda)$ is a light--like
vector and eq. (15) implies the Virasoro condition.

As in the $N=(1,0)$ case, the action consists of 3 terms:

$$
I=I^{(C)} + I^{(B)} + I^{(h)}.
$$

\noindent
$I^{(C)}$ imposes the twistor constraint. It is:

$$
I_c = \int d^2\xi d\eta d\bar\eta
\Bigl\{ \hat P_a \hat{\bar E}^a_1 - \hat{\bar P}_a
\hat E^a_1\Bigr\}\eqno(4.16)
$$

\noindent
where the lagrangian multipliers $\hat P_a$ are complex, w.s. superfields.

$I^{(B)}$ is the action term which involves
the 2-superform B.

According to our previous discussion, it can be written as

$$
I^{(B)} = \int_{{\cal M}_0} \tilde B\eqno(4.17)
$$

\noindent
where ${\cal M}_0$ is the slide of ${\cal M}$ at $\eta=0=d\eta$
and

$$
\tilde B = \hat B+e^+\wedge e^-
\hat E^A_1 \hat{\bar E}^B_1 E^C_+ \hat H_{CBA}=\hat B +
e^+ e^- \phi E^a_- E_{+a}
$$

\noindent
where the twistor and SUGRA--SYM constraints have been used.

\noindent
As a consequence of the Weil triviality of $\tilde B,I^{(B)}$ is
super--reparametrization invariant, even if it is not a full
superspace integral. Moreover

$$
I^{(B)} = \int_{{\cal M}_0} [e^+\wedge e^- \phi E^a_- E_{+a} + B]
\eqno(4.18)
$$

\noindent
coincides with the G.S. action (without the
heterotic fermions).

\noindent
$I^{(h)}$ is the heterotic  action. Consider a set of 16, complex,
covariantly chiral fermions $\hat\psi_r(r=1,...,16)$:

$$
{\cal D}_1 \hat\psi_r = 0 = \bar{\cal D}_1 \hat{\bar\psi}_r
$$

Then the heterotic action is

$$
I^{(h)} = i \int d^2\xi d\eta d\bar\eta \sum_r
\hat{\bar\psi}_r \hat{\psi}_r\quad
= i \int d^2\xi \sum_r
(\bar\psi_r {\cal D}_- \psi_r + \bar\chi_r \chi_r)
$$

\noindent
where $\chi_r = \bar{\cal D}_1 \hat\psi_r |_{\eta=0}$

Now the $\kappa$--symmetry has six superparameters since two
of the eight parameters in the G.S. approach,
here are absorbed by the $N=(2,0)$ supersymmetry.

As shown in [31],[32] the $N=(2,0)$ model is well suited to be quantized
and provides a consistent and useful, semicovariant quantization
scheme for the G.S. (heterotic) superstring. $\kappa$--symmetry
can be gauge fixed by killing six components of $\hat\theta^\alpha$.
More precisely, a spinor of $SO(1,9)$ can be decomposed under
$U(4)$ as follows

$$
16  \rightarrow 4\oplus 4 \oplus 1 \oplus 1 \oplus 6
$$

\noindent
and $\kappa$--symmetry is gauge fixed by gauging to zero the
$U(4)$ irrep. 6 of $\hat\theta^\alpha$.
$SO(1,9)$ is broken to $U(4)$ but the $N=(2,0)$ supersymmetry is
preserved and can be fixed in superconformal gauge.

The relevant point is that both the left and the right conformal
anomaly vanishes in this gauge. Let us recall that for a conformal
field with left (right) conformal weight $j_L(j_R)$ the left (right)
conformal anomaly is proportional to $\eta_L(n_R)$ where
$n=6j^2-6j+1$. For our model in this gauge, one has:

\vskip 0.5truecm

\def\3bar{\vrule width 1truecm height 0.01truecm}
\halign{\hskip 1.5truecm #\hfill & \hskip 0.5truecm \hfill#\hfill &
\hskip 0.5truecm \hfill#\hfill \cr
Field content: & $n_R:$ & $n_L:$ \cr
\noalign{\vskip 0.2truecm}
4 chiral and 4 antichiral bosonic superfields  \hfill
                                 & $4\times 3$ & $4\times 2$ \cr
\noalign{\vskip 0.2truecm}
2 chiral and 2 antichiral fermionic superfields \hfill
                                 & $-2\times 3$ &  +2         \cr
\noalign{\vskip 0.2truecm}
16 complex heterotic fermions \hfill   &   0          &  +16        \cr
\noalign{\vskip 0.2truecm}
1 reparametrization ghost     \hfill   & $-26$          & $-26$        \cr
\noalign{\vskip 0.2truecm}
2 supersymmetry ghosts        \hfill   & $+2\times 11$ &   0        \cr
\noalign{\vskip 0.2truecm}
1 vector ghost                \hfill   & $-2=$            &   $0=$        \cr
\noalign{\vskip 0.2truecm}
                              \hfill   & \3bar            &   \3bar        \cr
                              \hfill   & 0            &   0        \cr}

\vskip 0.5truecm

To appreciate this result one should notice that for the G.S.
heterotic string in semilight cone gauge [39], a simple calculation of
the right conformal anomaly, gives $n_R=10+4-26=-12$.
This ``anomaly"  turns out to be a trivial cocycle [40] in this gauge
so that it can be avoided perturbatively by subtracting suitable
counterterms. Therefore it does not implies an
inconsistency and no problems arise in $\sigma$--model perturbative
calculations.
However, due to this cocycle, higher stringly loop (i.e. higher
genus) calculations become impraticable. On the contrary, since
in the Berkovits quantization scheme of the $N=(2,0)$ model the
conformal anomaly vanishes, higher genus calculations of string
amplitudes become possible and in fact have been done successfully
[32].

\subsection{4.3}{D=10, N=(8,0), heterotic, twistor--string model
[28],[30]}

As in the previous cases, the action consists of 3 terms:

$$
I=I^{(C)} + I^{(B)} + I^{(h)}
$$

\noindent
$I^{(C)}$ impose the twistor constraint

$$
\hat E^a_q = 0\quad\quad q=1,...8\eqno(4.19)
$$

\noindent
and is given by

$$
I^{(C)} = \int d^2\xi d^8\eta\sum\limits^8_{q=1}
\hat P^q_a \hat E^a_q\eqno(4.20)
$$

\noindent
In components, the twistor constraint yields

$$
\lambda^a_q=0\eqno(4.21)
$$

$$
(\lambda_p\Gamma^a\lambda_q)=
\delta_{pq} E^a_-\eqno(4.22)
$$

\noindent
together with further constraints, irrelevant here. As for
$I^{(B)}$, it can be written in two equivalent ways in terms
of the modified two--superform [30]

$$
\tilde B = \hat B +
e^+\wedge e^- {1\over 8} \sum_q \hat E^A_q \hat E^B_q
\hat E^C_+\hat H_{CBA}
$$

\noindent
that, under the twistor and SUGRA--SYM constraints, satisfies eq. (5).

\noindent
They are: [28], [30]

$$
I^{(B)} = \int_{{\cal M}_0}\ d\tilde B,\eqno(4.23)
$$

\noindent
where ${\cal M}_0$ is the slide of ${\cal M}$ at $\eta^{(q)}=0=
d\eta^{(q)}$, or

$$
I'^{(B)} = \int d^2\xi d^8\eta \hat P^{IJ}
\Bigl( \hat B_{IJ} - \partial_I\hat Q_J\Bigr),\eqno(4.24)
$$

\noindent
where $\underline{\hat P}^{IJ}$ are Grassmann--antisymmetric,
w.s. superfields.

Under the twistor and SUGRA--SYM constraints, in both cases one
recover from eqs. (20) and (23) (or (24)) the G.S. action (without
heterotic fermions). Here the relevant point is that the action
(20) is invariant under a set of local abelian
transformations that involve $\hat P^q_a$ [30]. They are similar to
the Bianchi gauge transformations, eq. (8), and generalize the
$\beta$--symmetry
which is present in the $(1,0)$--model.
{}From
the Bianchi gauge transformations of $\hat P^{IJ}$
one can reduces
eq. (25) to eq. (24) and from these involving $\hat P^q_a$ one can
show that the lagrangian multipliers do not alter the field equations
of the G.S. approach. The same remark applies to the $N=(2,0)$
case.

The problem is with the
heterotic action $I^{(h)}$: unfortunately, up to now, nobody has
succeded to write an $N$--extended w.s. supersymmetric version of
$I^{(h)}$ for $N > (2,0)$ (in the proposal
of ref. [28], half of the heterotic fermions have negative norm).

\section{5}{Conclusions}
We have show that the introduction of commuting spinors (twistors)
leads to new formulations of superparticles and heterotic string
models that exhibit $n$--extended world sheet supersymmetry as well
as target space supersymmetry.

The twistor approach has  clarified the geometrical meaning of
$\kappa$--symmetry, which appears as (extended) world
sheet supersymmetry in disguise. Moreover it seems useful to clarify to some
extent, the relation between G.S. and N.S.R.  formulations.
At the quantum level, the $N=(2,0)$--model yields a new promising
method to calculate G.S. superstring amplitudes at higher stringly
loops.

There are, of course, open problems.

A consistent treatment of the heterotic fermions for $N>(2,0)$ is
still lacking.
Moreover it would be interesting to extend these twistor--like
formulations to supermembranes and non--heterotic superstrings.
In this direction work is in progress.

Finally the big problem is to
find new consistent quantization schemes for $N>(2,0)$ models, in
particular for the $D=10,N=(8,0)$ model.

\vskip 0.5truecm
\noindent
{\bf Acknowledgments.} I am grateful to N. Berkovits, P. Pasti and
D. Sorokin for stimulating discussions. This work has been partially
supported by the Ministero Italiano della Pubblica Instruzione.
\vskip 0.5truecm

\references
\paper{1}{A. Neveu and J.H. Schwarz}{Nucl. Phys}{B31}{1971}{86}
\vskip -0.2truecm
\item{\hbox to 20truept{}} {\rauth P. Ramond} {\it Phys. Rev.} {\bf D3}
(1971) 2415
\smallskip
\paper{2}{F. Gliozzi, J. Scherk and D. Olive}{Phys. Lett.}{65B}
{1976}{282}
\paper{3} {M.B. Green and J.H. Schwarz}{Phys. Lett.} {136B}
{1984}{367}
\paper{4} {W. Siegel} {Phys. Lett.} {128B} {1983} {397}
\paper{5} {R. Casalbuoni} {Nuovo Cim.} {33A} {1976} {310}
\paper{6} {E. Bergshoff, E. Sezgin and P.K. Townsend} {Phys. Lett.}
{189B} {1987} {75}
\paper{7} {E. Bergshoeff, R. Kallosh and A. Van Proeyen}{Phys. Lett.}
{251B}{1990}{128}
\vskip -0.2truecm
\item{\hbox to 20truept{}}{\rauth R. Kallosh} {\it Phys. Lett.} {\bf 251B}
(1990) 134
\vskip -0.2truecm
\item{\hbox to 20truept{}}{\rauth A. Mi\c covic, M. R\u ocek, W. Siegel,
P. Van Nieuwenhuizen, J. Yamron and A. Van de Ven} {\it Phys. Lett.}
{\bf 235B} (1990) 106
\paper{8} {A. Ferber}{Nucl. Phys.} {B132} {1978} {55}
\vskip -0.2truecm
\item{\hbox to 20truept{}} {\rauth T. Shirafuji} {\it Prog. Theor. Phys.}
{\bf 70} (1983) 18
\paper{9} {E. Witten} {Nucl. Phys.} {B266} {1986} {245}
\paper{10} {A. Bengtsson, I. Bengtsson, M. Cederwall and N. Linden} {Phys.
Rev.}
{D36} {1987} {1766}
\vskip -0.2truecm
\item{\hbox to 20truept{}} {\rauth I. Bengtsson, M. Cederwall}
{\it Nucl. Phys.} {\bf B302} (1988) 81
\paper{11} {Y. Eisemberg, S. Salomon} {Nucl. Phys.} {B309} {1988} {709}
\paper{12} {R. Penrose} {J. Math. Phys.} {8} {1967} {345}
\vskip -0.2truecm
\item{\hbox to 20truept{}} {\rauth R. Penrose and M. MacCallum}
{\it Phys. Reports} {\bf 66} (1972) 241
\paper{13} {D.P. Sorokin, V.I. Tkach and D.V. Volkov}
{Mod. Phys. Lett.} {A4} {1989} {901}
\paper{14} {D.P. Sorokin, V.I. Tkach, D.V. Volkov and A.A. Zheltukhin}
{Phys. Lett.} {216B} {1989} {302}
\vskip -0.2truecm
\item{\hbox to 20 truept{}} {\rauth D.P. Sorokin} {\it Fortschr.
Phys.} {\bf 38} (1990) 923
\vskip -0.2truecm
\item{\hbox to 20truept{}} {\rauth D.V. Volkov and A.A. Zheltukhin}
{\it Lett. Math. Phys.} {\bf 17} (1989) 141 and {\it Nucl. Phys.}
{\bf 335} (1990) 723
\vskip -0.2truecm
\item{\hbox to 20truept{}} {\rauth V.A. Soroka, D.P. Sorokin, V.I.
Tkach and D.V. Volkov} {\it Int. J. Mod. Phys.} {\bf 7A} (1992) 5977
\vskip -0.2truecm
\item{\hbox to 20truept{}} {\rauth A. Pashnev and D.P. Sorokin}.
{\it preprint JINR E2-92-27, Dubna}
\paper{15} {P.S. Howe and P.K. Townsend} {Phys. Lett.} {259B}
{1991} {285}
\paper{16} {P.K. Towsend} {Phys. Lett.} {261B} {1991} {65}
\paper{17} {J.P. Gauntlett} {Phys. Lett.} {272B} {1991} {25}
\vskip -0.2truecm
\item{\hbox to 20 truept{}} {\rauth Y. Eisenberg} {\it Phys. Lett.}
{276B} (1992) 325
\vskip -0.2truecm
\item{\hbox to 20truept{}} {\rauth M.S. Plyushchay} {\it Phys. Lett.}
{\bf 240B} (1991) 133
\paper{18} {F. Delduc and E. Sokatchev} {Class Quantum Grav.}
{9} {1992} {361}
\vskip -0.2truecm
\item{\hbox to 20 truept{}} {\it Phys. Lett.} {\bf 262B} (1991) 444
\paper{19} {A.S. Galperin, P.S. Howe and K.S. Stelle} {Nucl. Phys.}
{B368} {1992} {248}
\paper{20} {F. Delduc, A.S. Galperin and E. Sokatchev}
{Nucl. Phys.} {B368} {1992} {143}
\paper{21} {A.S. Galperin and E. Sokatchev} {Phys. Rev.} {D46}
{1992} {714}
\paper{22} {N. Berkovits} {Phys. Lett.} {232B} {1989} {184}
\paper{23} {N. Berkovits} {Nucl. Phys.} {B350} {1991} {193}
\vskip -0.2truecm
\item{\hbox to 20 truept{}} {\it Nucl. Phys.} {\bf B358} (1991) 169
\paper{24} {E.A. Ivanov and A.A. Kapustnikov} {Phys. Lett.}
{267B} {1991} {175}
\item{\hbox to 20truept{[25]}} {\rauth V. Chikalov and A. Pashnev}
{\it preprint JINR E2-92, Dubna}
\paper{26} {F. Delduc, E. Ivanov and E. Sokatcherv} {Nucl. Phys.}
{B384} {1992} {334}
\paper{27} {M. Tonin} {Phys. Lett.} {266B} {1991} {312}
\paper{28} {M. Tonin} {Int. J. Mod. Phys.} {7A} {1992} {6013}
\paper{29} {S. Aoyama, P. Pasti and M. Tonin} {Phys. Lett.}
{283B} {1992} {213}
\paper{30} {F. Delduc, A.S. Galperin, P. Howe and E. Sokatchev}
{preprint BONN-HE-92-19, JHU-TIPAC 920018, ENSLAM-L-392-92}{}{1992}{}
\paper{31} {N. Berkovits} {Nucl. Phys.} {B379} {1992} {96}
\item{\hbox to 20truept{[32]}} {\rauth N. Berkovits}
{\it Stony Brook preprint ITP-92-42} (1992);
\vskip -0.2truecm
\item{\hbox to 20truept{}} {\it King's College preprint KCL-TH-92-6} (1992)
\paper{33} {M. Grisaru, P. Howe, L. Mezincescu, B. Nilsson and
P.K. Townsend} {Phys. Lett.} {162B} {1985} {116}
\paper{34} {B.E.W. Nilsson} {Nucl. Phys.} {B188} {1981} {176}
\vskip -0.2truecm
\item{\hbox to 20truept{}} {\it Phys. Lett.} {\bf 175B} (1986) 319
\paper{35} {J.J. Atick, A. Dhar and B. Ratra} {Phys. Lett.}
{169B} {1986} {54}
\paper{36} {M. Tonin} {Int. J. Mod. Phys} {A3} {1988} {1519}
\vskip -0.2truecm
\item{\hbox to 20truept{}} {\bf A4} (1989); {\bf A6} (1991) 315
\paper{37} {L. Bonora, P. Pasti and M. Tonin} {Phys. Lett.}
{156B} {1986} {191;}
\vskip -0.2truecm
\item{\hbox to 20truept{}} {\it Nucl. Phys.} {\bf B261} (1985) 241;
{\bf B286} (1987) 150
\paper{38} {J.A. de Azcarraga, J.M. Izquierdo and P.K. Towsend}
{Phys. Rev.} {\bf D45} {1992} {3321}
\vskip -0.2truecm
\item{\hbox to 20truept{}} {\rauth P.K. Townsend} {\it Phys.
Lett.} {\bf 277B} (1992) 285
\paper{39} {S. Carlip} {Nucl. Phys.} {B284} {1987} {365}
\paper{40} {M. Porrati and P. Van Nieuwenhuizen} {Phys. Lett.}
{273B} {1991} {47}

\bye